\documentclass[11pt,epsf]{article}

\usepackage{epsfig}
\usepackage{amssymb}
\usepackage{graphicx}
\usepackage{color}
\usepackage{subfigure}

\begin{document}

\baselineskip 6mm
\renewcommand{\thefootnote}{\fnsymbol{footnote}}


\newcommand{\nc}{\newcommand}
\newcommand{\rnc}{\renewcommand}



\newcommand{\tcb}{\textcolor{blue}}
\newcommand{\tcr}{\textcolor{red}}
\newcommand{\tcg}{\textcolor{green}}


\def\be{\begin{equation}}
\def\ee{\end{equation}}
\def\ba{\begin{array}}
\def\ea{\end{array}}
\def\bea{\begin{eqnarray}}
\def\eea{\end{eqnarray}}
\def\nn{\nonumber\\}


\def\ct{\cite}
\def\la{\label}
\def\eq#1{(\ref{#1})}


\def\a{\alpha}
\def\b{\beta}
\def\g{\gamma}
\def\G{\Gamma}
\def\d{\delta}
\def\D{\Delta}
\def\ep{\epsilon}
\def\e{\eta}
\def\ph{\phi}
\def\Ph{\Phi}
\def\ps{\psi}
\def\Ps{\Psi}
\def\k{\kappa}
\def\l{\lambda}
\def\L{\Lambda}
\def\m{\mu}
\def\n{\nu}
\def\th{\theta}
\def\Th{\Theta}
\def\r{\rho}
\def\s{\sigma}
\def\S{\Sigma}
\def\ta{\tau}
\def\o{\omega}
\def\O{\Omega}
\def\pr{\prime}


\def\half{\frac{1}{2}}

\def\goto{\rightarrow}

\def\na{\nabla}
\def\grad{\nabla}
\def\curl{\nabla\times}
\def\div{\nabla\cdot}
\def\pa{\partial}

\def\bra{\left\langle}
\def\ket{\right\rangle}
\def\lb{\left[}
\def\lc{\left\{}
\def\ls{\left(}
\def\lp{\left.}
\def\rp{\right.}
\def\rb{\right]}
\def\rc{\right\}}
\def\rs{\right)}

\def\vac#1{\mid #1 \rangle}


\def\td#1{\tilde{#1}}
\def\check{ \maltese {\bf Check!}}


\def\Tr{{\rm Tr}\,}
\def\det{{\rm det}}


\def\bc#1{\nnindent {\bf $\bullet$ #1} \\ }
\def\ch {$<Check!>$ }
\def\ss {\vspace{1.5cm}}

\begin{titlepage}

\hfill\parbox{5cm} { }

\vspace{25mm}

\begin{center}
{\Large \bf Holographic $1/N_c$ correction from the chiral condensate }

\vskip 1. cm
  {Bum-Hoon Lee$^{ab}$\footnote{e-mail : bhl@sogang.ac.kr},
  Chanyong Park$^a$\footnote{e-mail : cyong21@sogang.ac.kr}
  and Sunyoung Shin$^a$\footnote{e-mail : sihnsy@gmail.com}}

\vskip 0.5cm

{\it $^a\,$ Center for Quantum Spacetime (CQUeST), Sogang University, Seoul 121-742, Korea}\\
{ \it $^b\,$ Department of Physics, Sogang University, Seoul 121-742, Korea}\\

\end{center}

\thispagestyle{empty}

\vskip2cm


\centerline{\bf ABSTRACT} \vskip 4mm

\vspace{1cm}
We investigate a gravity solution containing the gravitational backreaction of the massive
scalar field dual to the chiral condensate, which corresponds to $1/N_c$ correction. In general,
condensation changes the vacuum structure, so the present dual geometry is appropriate to
describe the chiral condensate vacuum in the gauge theory side.
After constructing the dual geometry numerically and applying the hard wall model
we study the effect of the $1/N_c$ correction on the lightest meson spectra,
which improves the values for lightest meson masses into the observations.
In addition, we investigate the chiral condensate dependence the binding energy
of heavy quarkonium.

\vspace{2cm}


\end{titlepage}

\renewcommand{\thefootnote}{\arabic{footnote}}
\setcounter{footnote}{0}

\tableofcontents

\section{Introduction}

Related to the RHIC and LHC experiments there have been many attempts
to understand the strongly interacting systems based on the AdS/CFT correspondence
\ct{Maldacena:1997re,Erdmenger:2007cm,Gursoy:2008za,Policastro:2002se,Herzog:2006ra,Sakai:2004cn,Erlich:2005qh,
Karch:2006pv,Da Rold:2005zs}. Recently, many holographic models describing the QCD-like
gauge theory in the strong coupling regime were invented. Moreover, various techniques used
in the holographic QCD are utilized in the holographic condensed matter system, which is called
AdS/CMT.

In the framework of a holographic QCD approach, the confinement was realized with an
infrared(IR) cut-off in the AdS space, so called the hard wall model \ct{Erlich:2005qh}. This hard wall
model can explain the masses of various mesons, comparing with observations,
within $10 \%$ error. In Ref. \ct{Herzog:2006ra} , by using this hard wall model the
deconfinement phase transition
was also investigated by identifying it with the Hawking-Page transition of the dual gravity theory.
There exists a different model, so called soft wall model \cite{Karch:2006pv}, in which by introducing a
non-dynamical scalar field the linear confinement behavior was explained. Usually, we call these
two models bottom-up approach. When we consider the string theory origin of the gravity theory,
there exist other approaches so called top-down approaches \cite{Sakai:2004cn},
in which after constructing some branes configuration
the dual gauge theory is investigated.

In this paper, we will concentrate on the hard wall model. In the original one \ct{Erlich:2005qh}, the $1/N_c$ corrections
was not fully considered. There are two different $1/N_c$ corrections. One is coming from the
medium effect composed of quarks in the fundamental representation of the gauge group
\cite{Da Rold:2005zs,Lee:2009bya}. Following the
AdS/CFT correspondence, the gravitational constant $1/2 \k^2$ is proportional to $N_c^2$ degrees of
freedom and the coupling constant of the dual bulk gauge field, which represents the
$SU(N_f)$ flavor group of the dual gauge theory, has $N_c N_f$ degrees of freedom. Since
the time-component of the bulk gauge field corresponds to the quark number density operator in the
dual gauge theory, considering the gravitation backreaction of this gauge field maps to
investigate the $1/N_c$ corrections coming from the quark density medium. After considering these
$1/N_c$ corrections and finding the dual geometries,
the Hawking-Page transition, various light meson spectra and the string breaking
of the heavy quarkonium were investigated.

There exists the other $1/N_c$ correction coming from the chiral condensate. In \cite{Kim:2007em},
the partial corrections of the chiral condensate without the gravitational backreaction were considered.
In \cite{Shock:2006gt}, the backreaction of the massive scalar field with a special type of scalar
potential was also investigated.
At present paper, the goal is to find a dual geometry
corresponding to the chiral condensate vacuum without introducing a special type of scalar potential
and investigate various meson spectra. Since  the chiral symmetry is restored
in the deconfining phase, we concentrate on the confining phase from now on. In the confining phase,
the chiral symmetry is spontaneously broken due to the chiral condensate, which was investigated
in the holographic model \cite{Kruczenski:2003uq}. Usually, this chiral condensate changes the
vacuum structure, which implies in the holographic QCD  that the dual
geometry needs to be deformed. Unlike the gluon condensate background where the analytic solution
including the gravitational backreaction of the gluon condensate was found \cite{Gubser:1999pk,Ko:2009jc},
in the chiral condensate background
it seems to be impossible to find the dual geometric solution analytically. In this paper, we will
calculated the dual geometry numerically and investigate various meson spectra. As will be shown
in Sec.4, in the dual geometry including full $1/N_c$ correction coming from the bulk massive
scalar field, which represents the chiral condensate and the light quark mass, we obtain
the light meson masses very similar to observations.

The rest parts follows: In Sec. 2, we construct the gravity theory dual to
chiral condensate vacuum, in which the massive scalar field in the bulk corresponds to the
quark mass and the chiral condensate, and find numerical solutions in the chiral limit.
Usually, this numerical solution has a naked singularity which causes the IR divergence
in the dual QCD, so we introduce a hard wall to avoid this IR divergence.
In Sec. 3, on this numerical background we investigate the light meson spectra and string breaking
of the heavy quarkonium in the chiral limit. In Sec. 4, we revisit the dual geometry when the
light quark mass is not zero. Due to the light quark mass, the asymptotic solution is totally
different with one in the chiral limit. After finding the full numerical solution, we investigate
the meson spectra depending on the chiral condensate. Especially, when the light quark mass $m_q$
and the chiral condensate $\s$ are $2.383$MeV and $(304 {\rm MeV})^3$ respectively, we obtained
a good result explaining the observations. Finally, in Sec. 5 we finish our work with some
remarks.

\section{Asymptotic AdS background with the chiral condensate}

Following the AdS/CFT correspondence, the ground state containing the chiral condensate
can be described by the dual gravity theory including the massive scalar field.
In this section, to find a dual geometry corresponding to the chiral condensate ground state
in QCD we start with the following action
\bea \la{orgact}
S &=& \int d^5 x \sqrt{-G} \lb \frac{1}{2 \k^2} \ls {\cal R} - 2 \L \rs \rp \nn
&& \qquad \qquad \lp - \Tr \lc | D \Ph |^2 +  m^2 |\Ph|^2 + \frac{1}{4 g^2} \ls F^{(L)}_{MN} F^{(L)MN} +
F^{(R)}_{MN} F^{(R)MN} \rs \rc \rb ,
\eea
where $m^2 = -\frac{3}{R^2}$ and the cosmological constant is given by $\L = -\frac{6}{R^2}$.
In the above, the superscripts, $(L)$ and $(R)$ in the last term imply the left and right part
of  $SU(N_f)_L \times SU(N_f)_R$ flavor
symmetry group with $F^{(L,R)}_{MN} = \pa_M A^{(L,R)}_N - \pa_N A^{(L,R)}_M
- i \lb A^{(L,R)}_M, A^{(L,R)}_N \rb$. In this paper, since we are interested in the
chiral condensate ground state we consider bulk gauge fields as fluctuations, which
correspond to various vector or axial-vector mesons.
In addition, $D$ means a covariant derivative, $D_M \Ph = \pa_M \Ph - i A^{(L)}_M  \Ph
+ i \Ph A^{(R)}_M $.
Under the flavor symmetry group, $SU(N_f)_L \times SU(N_f)_R$,
the complex scalar field $\Ph$ transforms as a bifundamental representation. Now, we set
\be
\Ph(z) = \frac{ \ph(z) \bf{1}  }{2 \sqrt{N_f}} \ e^{i \pi^a (z) T^a} ,
\ee
where $\bf{1}$ means an identity matrix and $T^a$ is a generator of the flavor symmetry group.
Furthermore, we identify $\ph(z)$ and $\pi^a (z)$ with the background field
and fluctuations respectively, in which the background field $\ph$ and fluctuation $\pi^a$
correspond to the quark mass (or chiral condensate) and pseudoscalar mesons.

From \eq{orgact}, the action describing only background fields reduces to
\be \la{backact}
S = \int d^5 x \sqrt{-G} \lb \frac{1}{2 \k^2} \ls {\cal R} - 2 \L \rs
- \frac{1}{4}  \lc \ls \pa \ph \rs^2 +  m^2 \ph^2 \rc  \rb .
\ee
The equations of motion for the metric and the scalar field $\ph(z)$ are given by
\bea  \la{eq:einstein}
{\cal R}_{MN} - \half G_{MN} {\cal R} + G_{MN} \L &=& \frac{\k^2}{2} \lb
\pa_M \ph \pa_N \ph - \half G_{MN} \ls ( \pa \ph )^2 + m^2 \ph^2 \rs \rb , \nn
0 &=& \frac{1}{\sqrt{G}} \pa_M \sqrt{G} G^{MN} \pa_N \ph - m^2 \ph .
\eea
Under the following ansatz
\be \la{metansmq0}
ds^2 = \frac{R^2}{z^2} \lb - F(z) dt^2 + G(z) d \vec{x}^2 + dz^2 \frac{}{} \rb ,
\ee
Einstein equation and the equation of motion for the scalar field become
\bea    \la{eq:full}
0 &=& \frac{F}{4 z^2 G} \lb \k^2 G \ls 3 \ph^2 - z^2 \ph'^2 \rs - 6 z \ls - 3 G' + z G''  \rs \rb , \nn
0 &=& \frac{1}{4}  \lb \ls \frac{2 F'}{F} - \frac{12}{z} \rs G' - \frac{G'^2}{G} +
\ls \k^2 \ph'^2 + \frac{2 z F'' - 6 F'}{z F} - \frac{3 \k^2 \ph^2}{z^2} - \frac{
F'^2}{F^2} \rs G + 4 G'' \rb, \nn
0 &=& \frac{1}{4 z^2 F G^2} \lb 3 z G F' \ls z G' -2 G  \rs
- F \ls 18 z G G' - 3 z^2 G'^2  + \k^2 G^2 (3 \ph^2 + z^2 \ph'^2 ) \rs \rb , \nn
0 &=& \frac{1}{4 R^2 F G}
\lb z^2 G F' \ph' + F  \ls 3 z^2 G' \ph' + 2 G (3 \ph - 3 z \ph'
+ z^2 \ph'' ) \rs \rb ,
\eea

Here, the solution we want is an asymptotic AdS space, so near the UV boundary ($z =0$)
the unknown functions $F$ and $G$ can be expanded to
\bea    \la{ans:metric}
F &=& 1 + \sum_{i=1}^{\infty} a_i z^i , \nn
G &=& 1 + \sum_{i=1}^{\infty} b_i z^i .
\eea
Notice that in the pure AdS background a scalar field with $m^2=-3/R^2$ has the following solution
\be
\ph = m_q z + \s z^3 ,
\ee
where $m_q$ and $\s=\bra \bar{q}{q} \ket$ are a current quark mass and chiral condensate, respectively.
From this fact,  when we consider the backreaction of the scalar field,
the solution of a massive scalar field near the UV boundary can be expanded as
\be \la{ans:scalar}
\ph = \s z^3 + \sum_{i=4}^{\infty} c_i z^i ,
\ee
where we set $m_q=0$, which corresponds to the chiral limit. If $m_q \ne 0$, the ansatz
in \eq{ans:metric}
and \eq{ans:scalar} should be modified to the form containing $\log z$ terms, see Sec. 4.
In this section, we concentrate on the chiral limit.

Near the UV boundary, the most general perturbative solutions up to ${\cal O} (z^{11})$ are given by
\bea    \la{sol:full}
F &=& 1 - 3 M z^4 - \frac{\k^2}{12} \s^2  z^6  + 4 M^2 z^8  + \frac{\k^2}{20}  M \s^2 z^{10} , \nn
G &=& 1 + M z^4 - \frac{\k^2}{12} \s^2  z^6 - \frac{ \k^2}{60} M \s^2 z^{10} , \nn
\ph &=& \s z^3  + \frac{ \k^2}{16} \s^3 z^9 ,
\eea
where new parameter $M$ implies the mass of the black hole. To see this, if turning off $\s$, the
above solutions reduce to the perturbative expansion form of the AdS black hole metric
in the Fefferman-Graham coordinate
\bea
F &=& \frac{(1 - M z^4)^2}{1 +M z^4} , \nn
G &=& 1 + M z^4 .
\eea

\begin{figure}
\begin{center}
\vspace{0cm}
\hspace{-1.cm}
\subfigure[]{ \includegraphics[angle=0,width=0.45\textwidth]{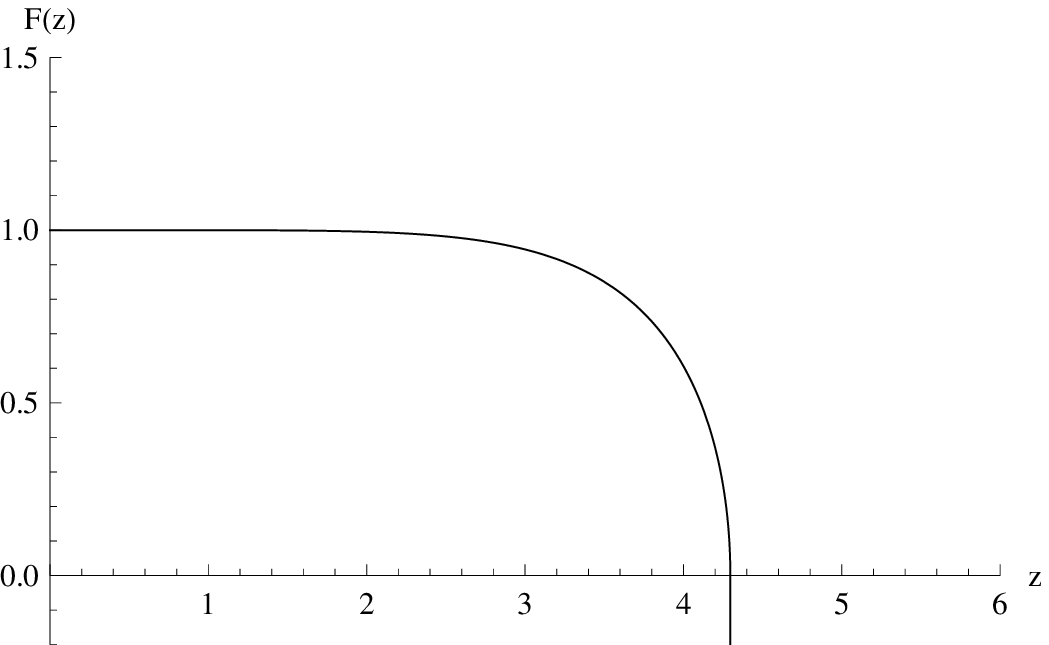}}
\hspace{0.5cm}
\subfigure[]{ \includegraphics[angle=0,width=0.45\textwidth]{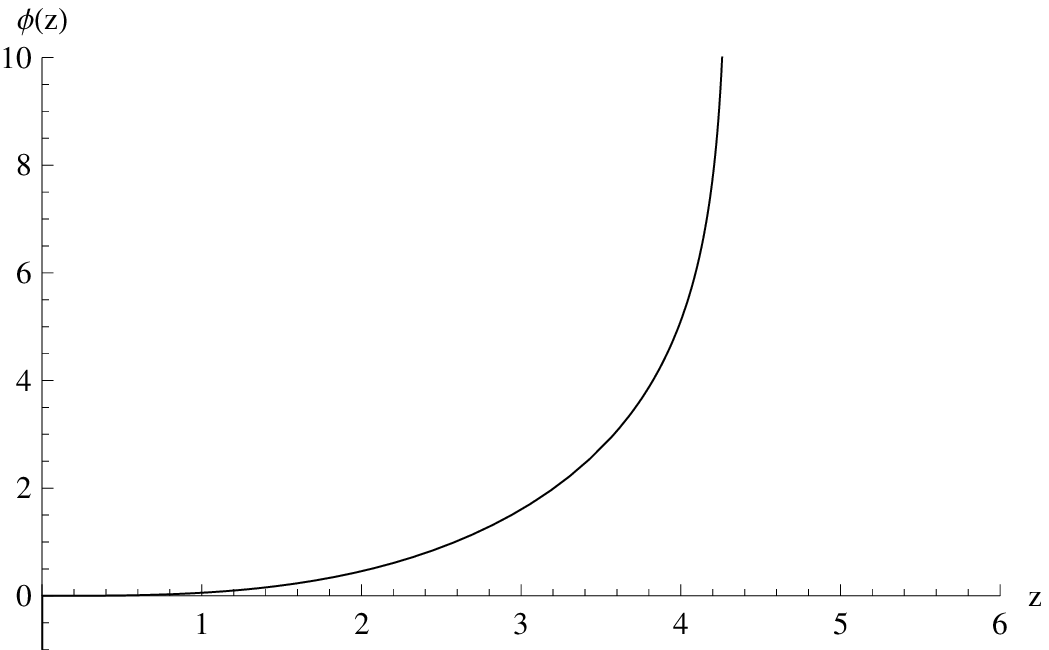}}
\vspace{0cm} \\
\caption{\small (a) The metric factor $F(z)$ (b) the scalar field $v(z)$ for $\s
= \ls 0.3846 {\rm GeV} \rs^3$, in which
the metric factor $F(z)$ becomes zero at $z=4.2963$. }
\label{number}
\end{center}
\end{figure}

In this paper, since we are interested in the meson spectra in the confining phase we
set $M=0$ with a non-zero chiral condensate. Notice that non-zero $M$ corresponds
to the Schwarzschild black hole solution dual to the deconfining phase.
From \eq{sol:full}, the perturbative solution describing
the chiral condensate background in the confining phase is given by
\bea    \la{sol:chiral}
F &=& G = 1 - \frac{\k^2}{12} \s^2  z^6  , \nn
\ph &=& \s z^3  + \frac{ \k^2}{16} \s^3 z^9 .
\eea
Here since the Lorentz symmetry on the boundary space is restored in the confining phase,
$F$ is the exactly same as $G$. The above in \eq{sol:chiral} is the solution of
\bea    \la{eq:chiral}
0 &=& \frac{1}{4 z^2} \lb \k^2 F \ls 3 \ph^2 - z^2 \ph'^2 \rs
- 6 z \ls z F'' -3 F' \rs \rb ,\nn
0 &=& \frac{1}{4} \lb  \frac{6 F'^2}{F^2} - \frac{3 \k^2 \ph^2}{z^2} - \frac{24 F'}{z F}
- \k^2 \ph'^2 \rb , \nn
0 &=& \frac{1}{2 R^2 F} \lb 2 z^2 F' \ph' + F \ls 3 \ph + z ( z \ph'' -3 \ph' ) \rs \rb ,
\eea
which can be obtained from \eq{eq:full} with $F(z)=G(z)$. Note that the first two equations
in \eq{eq:full} reduce to the same equation, the first in \eq{eq:chiral} and that
the second corresponds to a constraint equation.
Using the boundary behavior of the perturbative solution in \eq{sol:chiral}, we
can easily find the full solutions of \eq{eq:chiral} numerically.

For simplicity, we set $R=1$ and $N_c=3$ and use
$\frac{1}{\k^2} = \frac{\pi^2}{4 N_c^2}$ \cite{Erlich:2005qh}.
When the chiral condensate is given by $\s = \ls 0.327 {\rm GeV} \rs^3$, we solve the
equations in \eq{eq:chiral} numerically. As shown in the Figure 1, we find that there exists
a naked singularity at $z_s=4.2963$ (see Figure 2), where the metric factor $F(z)$ becomes zero.

\begin{figure}
\begin{center}
\vspace{1cm}
\subfigure{ \includegraphics[angle=0,width=0.6\textwidth]{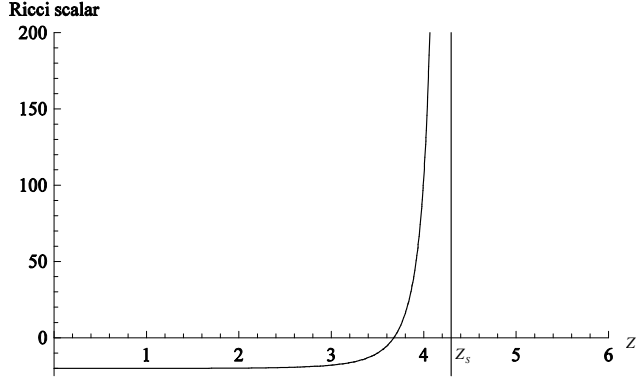}}
\vspace{-3cm}
\caption{\small  Ricci scalar for $\s = \ls 0.3846 {\rm GeV} \rs^3$, where the vertical
line implies the singular point.}
\label{number}
\end{center}
\end{figure}

The gravitational backreaction of a bulk massless scalar field corresponding to the
gluon condensate was investigated \cite{Ko:2009jc}. In the massless scalar case,
there exists a geometrical singularity, which can be considered as an IR cut-off.
Similarly, can we identify a geometrical singularity in the massive scalar case with an IR cut-off?
To answer this question, we should calculate the on-shell gravity action corresponding to the free
energy of boundary gauge theory. To do so,
we multiply the metric $G^{MN}$ to the first equation in \eq{eq:einstein}. Then, the Einstein
equation reduces to
\be
{\cal R} = \frac{10}{3} \L + \frac{\k^2}{2} \ls \pa \ph \rs^2 + \frac{5 \k^2}{6}  m^2 \ph^2 .
\ee
Inserting this to \eq{backact}, the on-shell action of this system becomes
\be
S_{on} = \int d^5 x \sqrt{-G} \lb \frac{2}{3 \k^2} \L + \frac{1}{6} m^2 \ph^2 \rb .
\ee
In the above, the geometrical singularity makes the on-shell gravity action diverge, so the
corresponding boundary free energy suffers from the IR divergence. As a result, it seems to be not
appropriate to consider the geometrical singularity as an IR cut-off in the massive scalar case.
One way to avoid
this problem is to introduce a hard wall as an IR cut-off in front of the geometrical
singularity $z_{IR} < z_s$.
From now on, we consider this hard wall approach only.

\section{Mesons in the chiral limit}

\subsection{light meson spectra}

Following the AdS/CFT correspondence, the bulk gauge field fluctuations correspond to the
vector mesons in the holographic QCD. In this section, we will investigate the spectrum of
the vector meson by turning on the vector part of the bulk gauge field, which is usually not mixed with
the scalar and axial gauge field. In the axial gauge $A_z=0$, the equation of motion
for the vector gauge field $V_i \equiv A^{(L)}_i + A^{(R)}_i $ ($i=1,2,3$) becomes
\be
0 = \frac{1}{\sqrt{G}} \pa_M \sqrt{G} G^{MP} G^{ii} \pa_P V_i  ,
\ee
where $M,N$ or $\m,\n$ are the five- or four-dimensional indices, respectively.
Using the ansatz,
\be
V_{\m} = \int \frac{d^4 k}{(2\pi)^4} \ e^{- i \o_n t + i \vec{p}_n \vec{x}} \  V^{(n)} (z) ,
\ee
where $n$ implies the $n$-th excitation mode, the above equation reduces to
\be     \la{eq:vector}
0 = \pa_z^2 V^{(n)} - \frac{F - z F' }{z F} \pa_z V^{(n)} + \frac{m_n^2}{F} V^{(n)} ,
\ee
where $m_n^2 = w_n^2 - \vec{p}_n^2$ is the mass square of the $n$-th excited meson state
and the prime means a derivative with respect to $z$. From now on, we will concentrate
on the lowest excited state, $\r$-meson, and denote its mass by $m_{\r}$.

Here, we have two initial parameters, $\s$ and $z_{IR}$, so that we should fix these values
to evaluate the meson masses.
As will be shown, since there is no direct interaction between the bulk gauge field and
the scalar field the vector meson mass does not strongly depend on the chiral condensate.
So we can choose $z_{IR} = 1/(0.3227 {\rm GeV})$, which was used in
the Ref. \cite{Erlich:2005qh} and gave good results for meson masses.
Notice that
the chiral condensate affects indirectly on the vector meson mass through the change of
the metric, though it is very small.

For the axial vector meson $A_{\m} \equiv A^{(L)}_{\m} - A^{(R)}_{\m} $ and pseudoscalar meson $\pi^a$,
we use the following ansatz
\bea
A_z &=& 0 , \nn A^a_{\m} &=& \int d^4 k \ e^{i q x} \ls \bar{A}^{a}_{\m} + \pa_{\m} \chi^a \rs , \nn
\pi^a &=& \int d^4 k \ e^{i q x} \pi^a ,
\eea
where $\bar{A}$ means a transverse gauge field in the bulk,
satisfying $\pa_{\m} \bar{A}^{\m} = 0$. Then, the equations of motion for
them are reduce to
\bea    \la{eq:axial}
0 &=& \pa_z \ls \frac{F}{z} \pa_z \bar{A}_{\m} \rs + \frac{m_{a_1}^2}{z} \bar{A}_{\m}
      -  \frac{g_5^2 F \ph^2}{z^3}  \bar{A}_{\m} ,\nn
0 &=& \pa_z \ls \frac{F}{z} \pa_z \chi \rs +
       \frac{ g_5^2 F \ph^2}{z^3}  \ls \pi - \chi \rs, \nn
0 &=& - m_{\pi}^2 \pa_z \chi + \frac{ g_5^2 F \ph^2}{z^2}  \pa_z \pi ,
\eea
where $m_{a_1}$ and $m_{\pi}$ are masses of the lowest excited mode
in the axial vector and pseudoscalar meson, respectively.
Moreover, using the holographic recipe \cite{Erlich:2005qh}, the pion decay constant $f_{\pi}$
is given by
\be
f_{\pi}^2 = - \lp \frac{1}{g_5^2} \frac{\pa_z \bar{A}(0)}{z} \right|_{z=0} ,
\ee
where $\bar{A}(0)$ is a solution of the first equation in \eq{eq:axial} at $m_{a_1}=0$,
satisfying two boundary conditions $\pa_z \bar{A} (z_{IR}) = 0$ and $\bar{A} (0) =1$.

To see the chiral condensate effect on the meson spectra, we should numerically solve
\eq{eq:vector} and \eq{eq:axial} with appropriate boundary conditions,
Neumann and Dirichlet boundary condition at $z=z_{IR}$ and $z=0$ respectively.
Before doing that, we should notice that
if $\s$ is greater than $\s_c = \ls 0.5337 {\rm GeV} \rs^3$,
the singular point $z_s$ is smaller than
$z_{IR}$. So the on-shell gravity action diverges, which implies that there exists an upper bound
of the chiral condensate value. From now on, we consider the case $\s < \s_c$ only.
In the Table 1, we numerically calculate the first-excited meson masses depending on the chiral condensate
and we plot these data in Figure 3. As the chiral condensate
decreases, the vector meson mass grows slightly but
the axial-vector meson mass, the pion mass, and the pion decay constant decrease.
Moreover, the masses of the vector and axial-vector meson have the same value at $\s=0$, which is due to
the restoration of the flavor symmetry.  In addition, since the pion mass is sufficiently small
the Gell-Mann-Oakes-Renner (GOR)
relation is `weakly' satisfied up to $10^{-12}$GeV$^4$ order in the chiral limit $m_q = 0$
\be
f_\pi^2 m_\pi^2 \approx 2 m_q \s .
\ee
In Figure 4, we plot the value of $\D = f_\pi^2 m_\pi^2$ and the chiral condensate dependence of
the pion decay constant.

\begin{table}
\begin{center}
\begin{tabular}{|c||c|c|c|c|c|c|}
\hline
$\s $ (GeV$^3$) & $m_{\rho}$ (GeV) & $m_{a_1}$ (GeV) & $m_\pi$ (GeV)         & $f_\pi$(GeV)               &$\Delta$ (GeV$^4$)              \\
\hline \hline
$0$             & $0.7760$         & $0.7760$    &                       &                       &                      \\
\hline
$(0.100)^3$     & $0.7760$         & $0.7769$    & $2.95\times10^{-6}$   & $4.79\times10^{-3}$   & $1.99\times10^{-16}$ \\
\hline
$(0.200)^3$     & $0.7760$         & $0.8304$    & $3.15\times10^{-6}$   & $3.59\times10^{-2}$   & $1.28\times10^{-14}$ \\
\hline
$(0.250)^3$     & $0.7760$         & $0.9598$    & $3.66\times10^{-6}$   & $6.03\times10^{-2}$   & $4.88\times10^{-14}$ \\
\hline
$(0.304)^3$     & $0.7758$         & $1.2217$    & $4.78\times10^{-6}$   & $8.31\times10^{-2}$   & $1.58\times10^{-13}$ \\
\hline
$(0.350)^3$     & $0.7755$         & $1.4754$    & $6.18\times10^{-6}$   & $9.81\times10^{-2}$   & $3.68\times10^{-13}$ \\
\hline
$(0.385)^3$     & $0.7751$         & $1.6442$    & $7.46\times10^{-6}$   & $10.82\times10^{-2}$  & $6.51\times10^{-13}$ \\
\hline
$(0.400)^3$     & $0.7748$         & $1.7115$    & $8.04\times10^{-6}$   & $11.25\times10^{-2}$  & $8.19\times10^{-13}$ \\
\hline
$(0.450)^3$     & $0.7735$         & $1.9281$    & $10.18\times10^{-6}$  & $12.66\times10^{-2}$  & $1.66\times10^{-12}$ \\
\hline
$(0.500)^3$     & $0.7709$         & $2.1424$    & $12.57\times10^{-6}$  & $14.07\times10^{-2}$  & $3.12\times10^{-12}$ \\
\hline
\end{tabular}
\end{center}
\caption{Various meson masses depending on the chiral condensate.
Here, the subscripts in the first line, $\r$, $a_1$ and $\pi$, imply the first excited
modes of the vector, axial vector and pseudoscalar mesons.}
\end{table}

\subsection{binding energy of the heavy quarkonium}

In this section, we will investigate the binding energy of a heavy quarkonium depending on the
chiral condensate.
First, we consider a heavy quarkonium composed of two heavy quarks on the chiral condensate
background obtained by numerical calculation in the previous section.
Then, the action for a fundamental string connecting two heavy quarks is given by
\be
S = \frac{1}{2 \pi \a'} \int d^2 \s \ \sqrt{\det \ h_{\a \b}} ,
\ee
where $h_{\a \b}$ is an induced metric on the string worldsheet.
For simplicity, we take $R=1$ and $\a' = 1/2 \pi$.

\begin{figure}[h!]
\vspace{0cm}
\begin{center}
 $
  \begin{array}{ccc}
  \epsfxsize=7.5cm
  \epsfbox{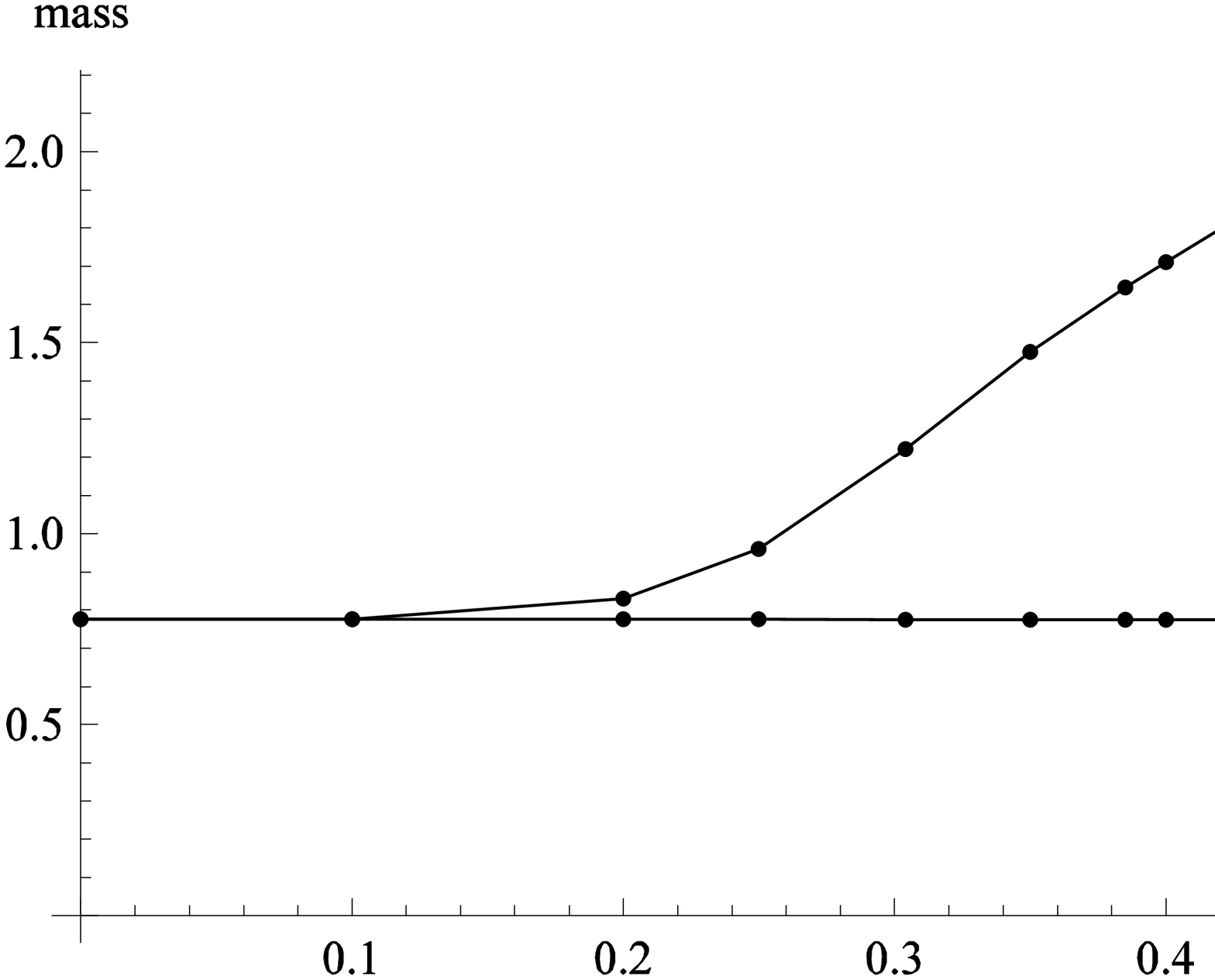} &
                           &
  \epsfxsize=7.5cm
   \epsfbox{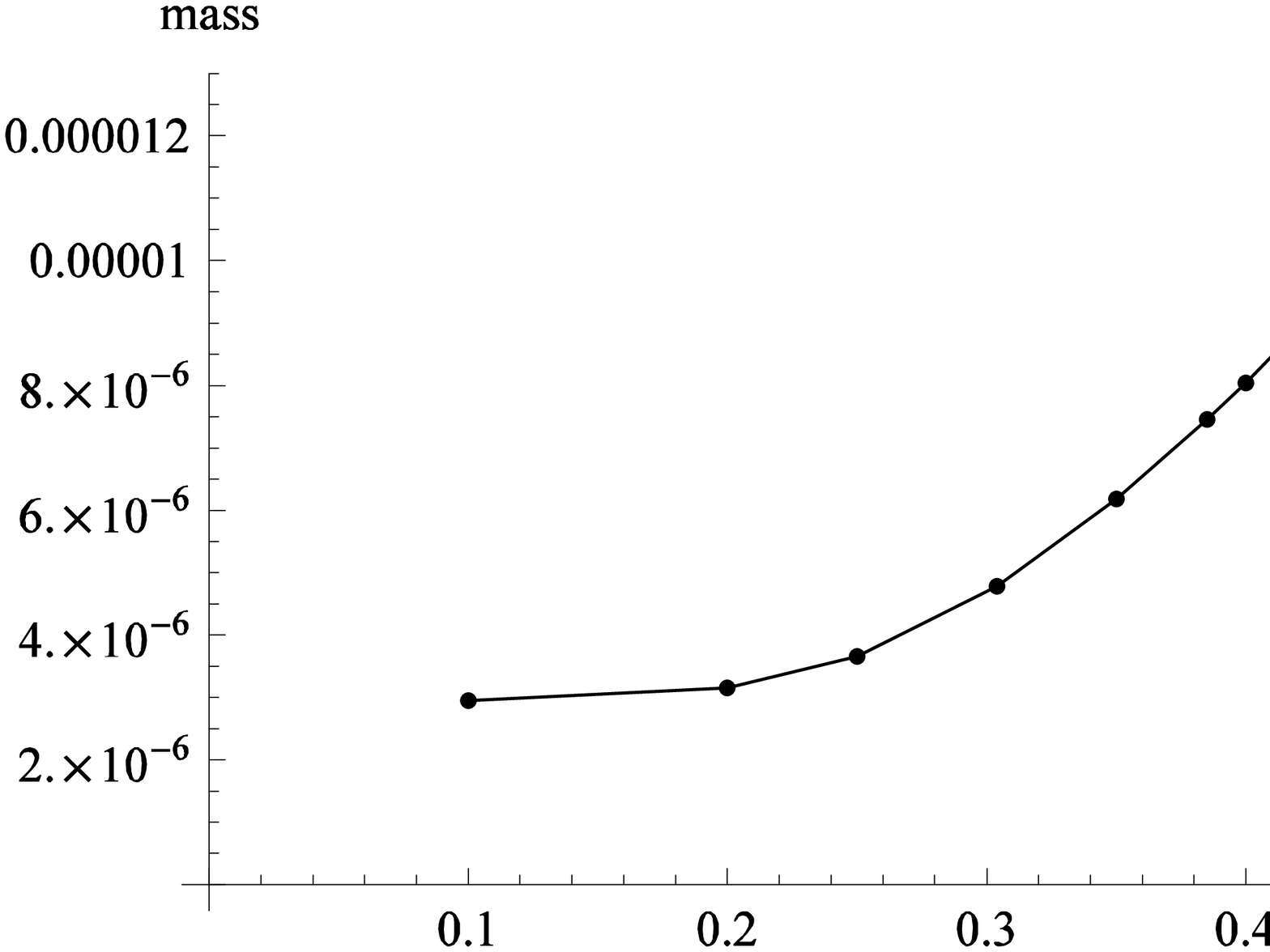}
  \end{array}
 $  \\
 $
  \begin{array}{ccc}
  \epsfxsize=7.5cm
   \epsfbox{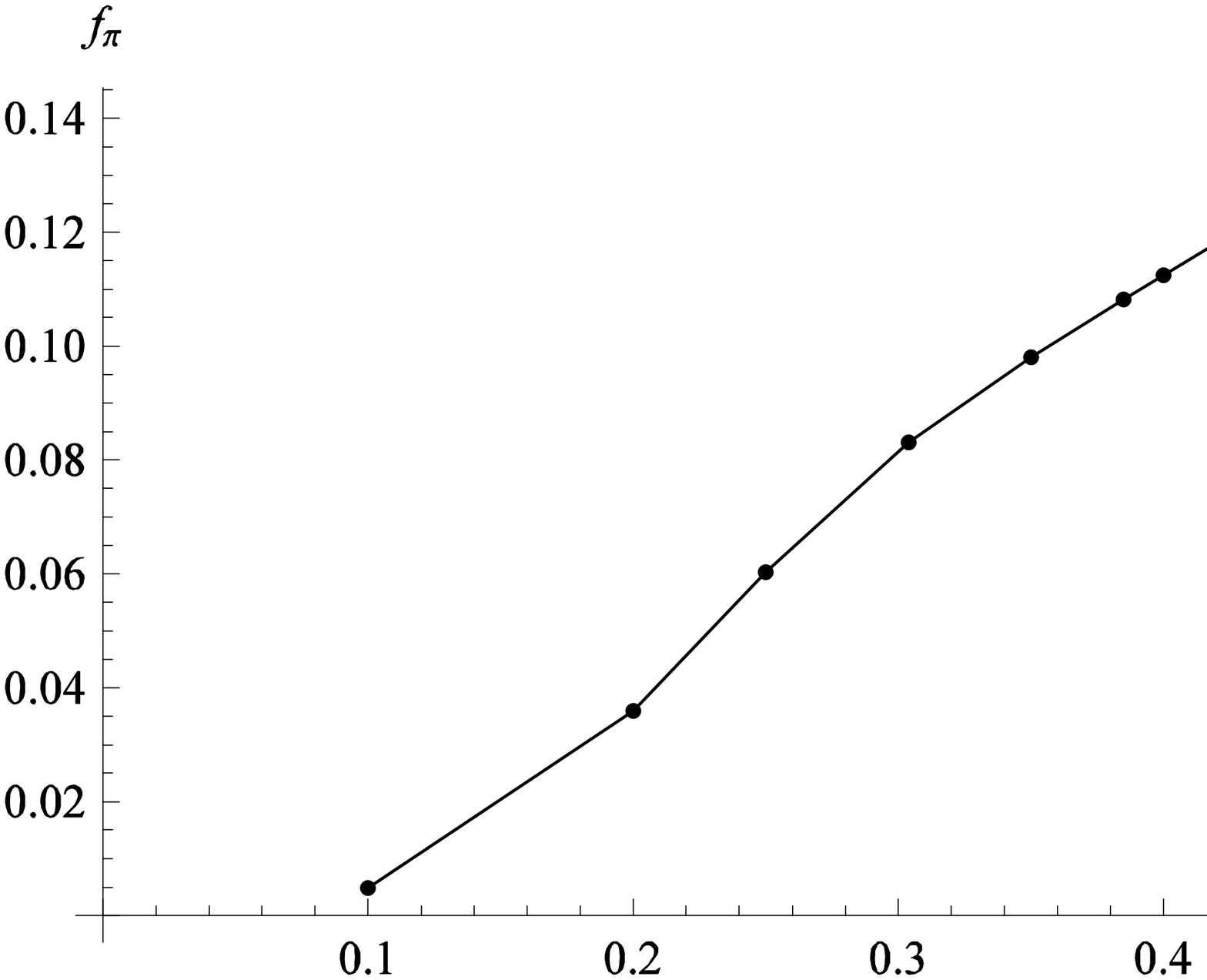} &
                           &
  \epsfxsize=7.5cm
   \epsfbox{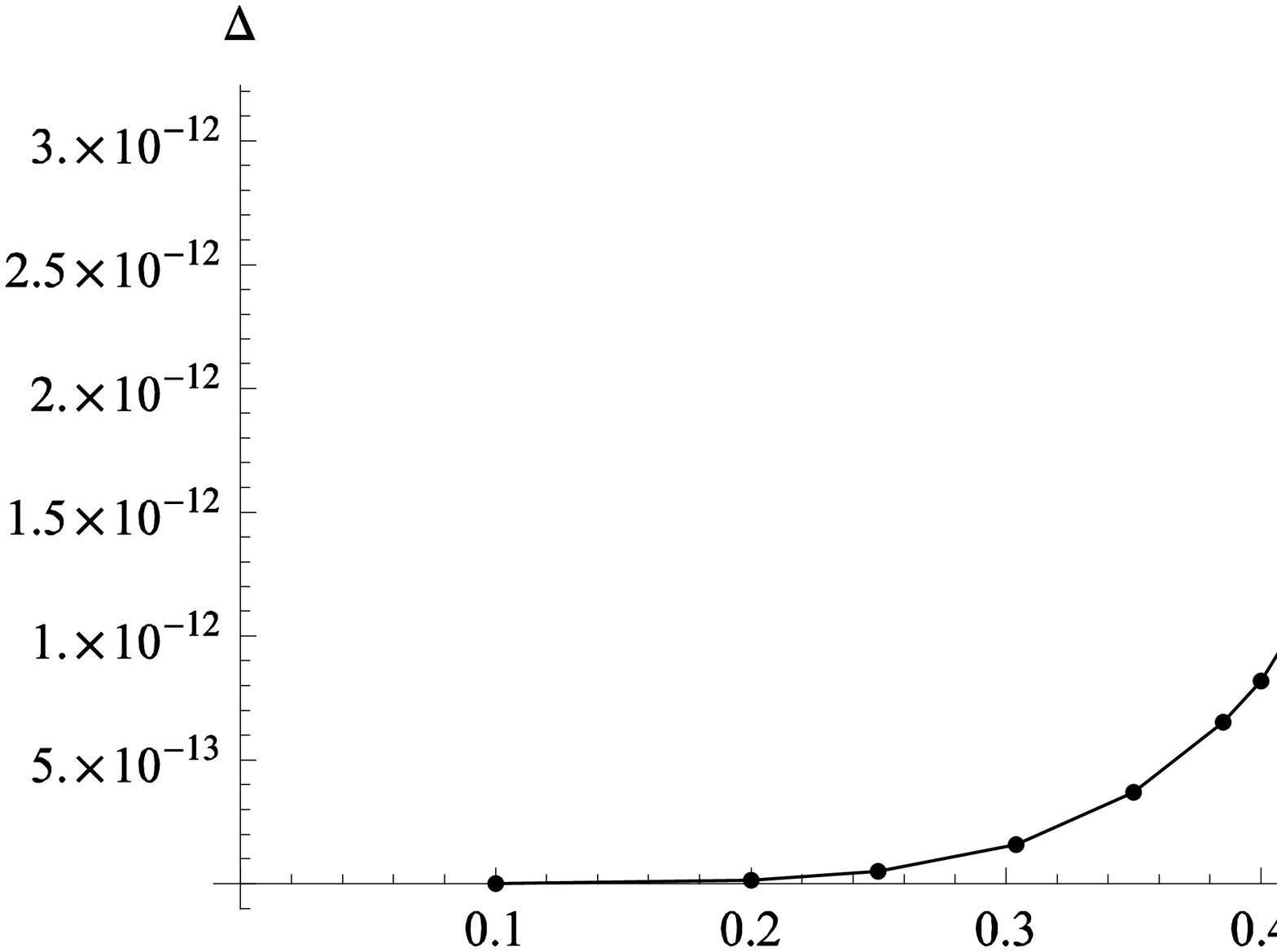}
  \end{array}
 $
 \vspace{-2.5cm}
\caption{Chiral condensate dependence of vector-, axial vector-mesons, pion and the pion decay
constant. In the right-below figure, we plot the deviation from GOR relation, in which
we found that GOR relation is almost satisfied in the chiral limit.}
\label{0mq_f_delta}
\end{center}
\end{figure}

Under the following gauge fixing
\be
\ta=t, \quad \sigma=x^1 \equiv x  \quad {\rm and} ~~ z=z(x) ,
\ee
the fundamental string action on the background metric \eq{metansmq0} with $F=G$ reduces to
\be
S= \int_{-T/2}^{T/2} dt \int_{-r/2}^{r/2}
dx \   \frac{\sqrt{F^2 + F {z^\prime}^2}}{z^2} \, ,
\ee
where $r$ is the distance between quark and anti-quark in the $x^1$ direction.
The conserved Hamiltonian of this system is
\be \label{ham1}
H = - \frac{F^2}{z^2 \sqrt{F^2 + F {z^\prime}^2}} .
\ee
If there exists $z_0$ where $\left. \frac{\pa z}{\pa x} \right|_{z=z_0}$ becomes zero,
the Hamiltonian at this point is given by
\be \la{eq2}
H = -  \frac{F_0}{z_0^2} ,
\ee
where $F_0$ is the value of $F$ at $z=z_0$. Due to the conservation law, we can find
the integral relation between $r$ and $z_0$ from two equations, \eq{ham1} and \eq{eq2}
\be
r=2  \int_0^{z_0} dz  \ \frac{z^2 F_0}{\sqrt{F \ls z_0^4 F^2 - z^4 F_0^2 \rs}} ,
\ee
and the energy of this string configuration
\bea
E &=&
2 \int_0^{z_0} dz \ \frac{z_0^2 F^{3/2}}{z^2 \sqrt{z_0^4 F^2 - z^4 F_0^2 }} .
\eea
Notice that since we are considering the static string configuration the kinetic energy
of two heavy quarks can be ignored. Then, the above energy corresponds to the potential energy
between two heavy quarks including the heavy quark masses.
Notice that the above potential energy diverges because in the hard wall model the
heavy quark have infinite mass. To consider the interaction energy only, we should subtract
this infinite mass term described by the straight string configuration.
To do so, we consider two straight strings describing the infinite mass of two quarks with the following
ansatz
\be
\ta=t \quad {\rm and} \quad \sigma=z .
\ee
Then, the energy of two straight strings become
\bea
2 M &=& 2 \int_0^{z_{IR}} dz \ \frac{\sqrt{F} }{z^2} ,
\eea
which diverges and can be interpreted as masses of two heavy quarks.
So the interaction energy of heavy quarkonium $V$ is finally given by
\bea    \la{intenergy}
V &=&
2 \int_0^{z_0} dz \ \frac{z_0^2 F^{3/2}}{z^2 \sqrt{z_0^4 F^2 - z^4 F_0^2 }}
 - 2 \int_0^{z_{IR}} dz \ \frac{\sqrt{F} }{z^2} ,
\eea
which is finite. If the interaction energy between two heavy quarks is larger than the mass
of two light quarks, the heavy quarkonium can break into two heavy-light meson, which is a
bound state of the heavy and light quarks. Usually, this phenomena is called the string breaking.
Here, since we assume that the mass of the light quark is zero,
the string breaking occurs when the interaction energy becomes zero. In Figure 4, we plot the
string breaking distance, at which the heavy quarkonium breaks to two heavy-light mesons.
As the chiral condensate increases the string breaking distance becomes short, which implies
that it is easy to break the heavy quarkonium to two heavy-light meson at the small
chiral condensate regime.


\begin{figure}
\vspace{0cm}
\begin{center}
   \epsfxsize=7.5cm
   \epsfbox{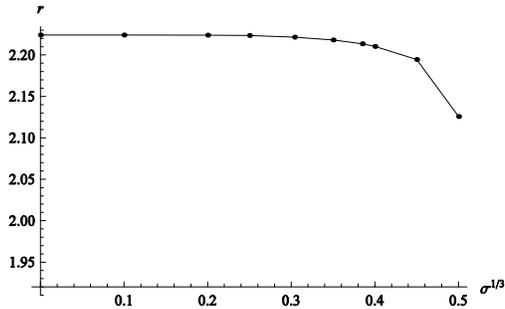}
     \vspace{-2.5cm}
  \caption{In the chiral limit $m_q=0$, the string breaking distance depending on the chiral condensate. }
\label{dissociation length mq=0}
\end{center}
\end{figure}

\section{Mesons with non-zero $m_q$}

\subsection{light meson spectra}

In this section, we improve the hard wall model to the case with non-zero quark mass.
As mentioned previously, the metric solution of \eq{eq:einstein} can be described by
a most general ansatz
\be
ds^2 = \frac{R^2}{z^2} \lb - F(z) dt^2 + G(z) d \vec{x}^2 + dz^2 \frac{}{} \rb .
\ee
This ansatz corresponds to the scalar field deformation of the black hole solution in the
Fefferman-Graham coordinate, see an example \eq{sol:full}. So, this ansatz is proper to describe
the black hole geometry dual to the deconfinement phase. For the confining phase,
the dual geometry should be described by a deformation of the AdS type metric, where the Lorentz
symmetry of the boundary space is restored. Therefore, the good ansatz for the dual geometry
of the confining phase is
\be
ds^2 = \frac{R^2}{z^2} \lb F(z) \ls - dt^2 + d \vec{x}^2 \rs  + dz^2 \frac{}{} \rb .
\ee
Since our interest is to investigate the meson spectra in the confining phase including
the chiral condensate, we concentrate on the latter case. Then, the equations of motion
for this system becomes \eq{eq:chiral}. Note that to find
the perturbative solutions of \eq{eq:chiral}, the appropriate $\log$ terms should be
added to the ansatz for metric and the scalar field. With those log terms,
the perturbative solutions near the boundary are given by
\bea    \la{fmqnzero}
F(z)&=&1 - \frac{\k^2}{12}   m_q^2 z^2 +
 \frac{1}{144} \ls \frac{}{} \k^4 m_q^4 - 18 \k^2 m_q \s - 3 \k^4 m_q^4 \log z \rs z^4 \nn
 && + \frac{1}{31104}
 \ls \frac{}{} 65 \k^6 m_q^6 - 396 \k^4 m_q^3 \s - 2592 \k^2 \s^2 -
    66 \k^6 m_q^6 \log z \right. \nn
    && \left. - 864 \k^4 m_q^3 \s \log z -
    72 \k^6 m_q^6 \log z^2 \frac{}{} \rs z^6  + \cdots
\eea
\bea
\ph (z) &=& m_q z + \ls \s + \frac{1}{6} \k^2 m_q^3 \log z \rs z^3  +
 \frac{1}{576}  \ls \frac{}{} -13 \k^4 m_q^5 + 144 \k^2 m_q^2 \s +
    24 \k^4 m_q^5 \log z \rs z^5 \nn
    && + \frac{1}{62208 } \ls \frac{}{}
  -209 \k^6 m_q^7 + 1044 \k^4 m_q^4 \s +
    10368 \k^2 m_q \s^2 + 174 \k^6 m_q^7 \log z  \right. \nn
  &&  \quad  \quad \quad \quad \quad \left. + 3456 \k^4 m_q^4 \s \log z +
    288 \k^6 m_q^7 \log z^2 \frac{}{} \rs z^7\nn
  &&   + \frac{1}{23887872}  \ls -2171 \k^8 m_q^9 -
    165744 \k^6 m_q^6 \s + 1586304 \k^4 m_q^3 \s^2 +
    1492992 \k^2 \s^3  \frac{}{} \right. \nn
    &&  \quad  \quad \quad \quad \quad \left. \frac{}{} - 27624 \k^8 m_q^9 \log z +
    528768 \k^6 m_q^6 \s \log z +
    746496 \k^4 m_q^3 \s^2 \log z  \right. \nn
    &&  \quad  \quad \quad \quad \quad \left. \frac{}{} + 44064 \k^8 m_q^9 \log z^2 +
    124416 \k^6 m_q^6 \s \log z^2 + 6912 \k^8 m_q^9 \log z^3 \rs z^9 \nn
    && + \cdots .
\eea
This solution reduces to \eq{sol:chiral} when $m_q = 0$. Using these as initial data for solving
the equations of motion, we can find various meson spectra depending on the chiral condensate
numerically, see Table 2 and Figure 5. Comparing Table 2 with Table 1, we
can see that though the masses of the vector and axial-vector meson
do not crucially depend on the
current quark mass, the magnitude of the current quark mass is very important to determine
the pion mass and pion decay constant. Here, for more realistic data,
we choose $z_{IR} = 1/(0.3227$GeV) and $m_q = 0.002383$GeV. In this case,
which give good data similar to the observed meson spectra.
In the improved EKSS model including the $1/N_c$ correction of the chiral condensate,
the masses of $\r$-, $a_1$-meson and pion at $\s = (304 {\rm MeV})^3$ are almost the same as the
experimental data (see Table 2). In Table 3, we compare our results with experimental observations
\cite{Eidelman:2004wy} and
those in the EKSS B-model \cite{Erlich:2005qh}.
Furthermore, we plot data of Table 2 in Figure 5.

\begin{table}
\begin{center}
\begin{tabular}{|c||c|c|c|c|c|c|}
\hline
 $\s $ (GeV$^3$) & $m_{\r}$ (GeV) & $m_{a_1}$ (GeV) & $m_\pi$ (GeV) & $f_\pi$ (GeV)  & $\Delta$ (GeV$^4$)\\
\hline \hline
$0$          & $0.77604$   & $0.7762$     & $0.7762$ & $0.0084$ & $4.30\times 10^{-5}$\\
\hline
$(0.100)^3$  & $0.77604$   & $0.7777$     & $0.3946$ & $0.0108$ & $1.35\times 10^{-5}$\\
\hline
$(0.150)^3$  & $0.77604$   & $0.7884$     & $0.2400$ & $0.0200$ & $6.99\times10^{-6}$\\
\hline
$(0.200)^3$  & $0.77602$   & $0.8354$     & $0.1698$ & $0.0388$ & $5.19\times10^{-6}$\\
\hline
$(0.250)^3$  & $0.77596$   & $0.9675$     & $0.1427$ & $0.0624$ & $4.97\times10^{-6}$\\
\hline
$(0.304)^3$  & $0.77580$   & $1.2306$     & $0.1396$ & $0.0846$ & $5.75\times10^{-6}$\\
\hline
$(0.350)^3$  & $0.77549$   & $1.4834$     & $0.1462$ & $0.0995$ & $7.06\times10^{-6}$\\
\hline
$(0.385)^3$  & $0.77506$   & $1.6517$     & $0.1528$ & $0.1096$ & $8.32\times10^{-6}$\\
\hline
$(0.400)^3$  & $0.77480$   & $1.7189$     & $0.1556$ & $0.1138$ & $8.89\times10^{-6}$\\
\hline
$(0.450)^3$  & $0.77345$   & $1.9354$     & $0.1650$ & $0.1279$ & $1.10\times 10^{-5}$\\
\hline
$(0.500)^3$  & $0.77083$   & $2.1497$     & $0.1739$ & $0.1419$ & $1.34\times 10^{-5}$\\
\hline
\end{tabular}
\caption{meson masses with the chiral condensate in the hard wall approach. ($\Delta=f_\pi^2m_\pi^2-2m_q\sigma$)}
\end{center}
\end{table}

There are several remarkable points.

\noindent {\bf 1.} Notice that although there is no direct coupling between the vector meson
and the massive scalar field in this model,
due to the gravitational backreaction of the massive scalar field the $\r$-meson
mass depends on the massive scalar field indirectly. From this fact, we can expect that
the $\r$-meson mass does not significantly depend on the massive scalar field
corresponding the chiral condensate. In Figure 3 and Figure 5, we can see that increasing of
the chiral condensate does not significantly change the $\r$-meson mass. Instead,
the $\r$-meson mass grows down slightly as the chiral condensate increases.

\noindent {\bf 2.} When we turn on the quark mass and chiral condensate,
the improved EKSS model including the chiral condensate gives good results
consistent with the measured values (see
the case with $m_q = 2.383$MeV and $\s= \ls 304  {\rm MeV}\rs^3$ in Table 2 and Table 3).

\noindent {\bf 3.} In the chiral limit, because the Lorentz symmetry of the boundary thoery
is fully restored
at $\s=0$, $\r$- and $a_1$-meson has the same mass, which was also studied in Ref. \cite{Kim:2008hx}
without considering the gravitational backreaction of the scalar field.
For the non-zero quark mass case, the chiral symmetry is almost restored in the limit
$\s \to 0$ although the explicit quark mass breaks this chiral symmetry slightly.
So, at $\s=0$ in Table 2, the $\r$-meson mass is slightly different with those of
$a_1$-meson and pion.

\noindent {\bf 4.} In Figure 5 and Table 2, there exists a critical value of the chiral condensate
$\s \approx \ls 0.26  {\rm GeV}\rs^3$.
As the chiral condensate decreases, the pion mass grows down above this critical point as
we expect, while below it the pion mass grows. This non-trivial chiral condensate dependence of
the pion mass can be understood in the gravity side by the flavor symmetry restoration and
the mixing between the pion and the longitudinal mode of the axial vector. Due to the flavor
symmetry restoration at the small $\s$ region, all vector and axial vector mesons have the
similar mass. Furthermore, for $m_q \ne 0$ the mixing between the pion and the longitudinal mode
of the axial vector
make the pion mass have the similar mass.
In the chiral limit as shown in Figure3, there is no such chiral condensate
dependence of the pion mass. In this case although the flavor symmetry is fully restored
the mixing between the pion and the longitudinal mode of the axial vector disappears
because $\ph$ becomes zero when $\s \to 0$. This fact implies that the small current quark mass
plays an important role to determine the pion mass in the small chiral condensate regime.
Though we do not have good understanding about the meaning of this behavior in the dual QCD side,
since the ratio between the pion mass and the chiral condensate $m_{\pi}/\s^{1/3}$ is larger than $1$
below the critical chiral condensate,
the usual chiral perturbation method in QCD does not work in this regime.
As a result, the results obtained here would shed light on understanding
physics, which can not be described by the chiral perturbation theory.

\begin{table}
\begin{center}
\begin{tabular}{|c||c|c|c|c|c|c|}
\hline
             & Measured \cite{Eidelman:2004wy} & EKSS B-model \cite{Erlich:2005qh} & chiral condensate background \\
\hline \hline
$z_{IR}$       &   & $1/(346 \ {\rm MeV})$     & $1/(322.7 \ {\rm MeV})$ \\
\hline
$m_q$          &   & $2.3$ MeV     & $2.383$ MeV \\
\hline
$\s$           &     & $ \ls 308 \ {\rm MeV} \rs^3$      & $ \ls 304 \ {\rm MeV} \rs^3$  \\
\hline
$m_{\r}$  & $775.8 \pm 0.5$  MeV   & $832$  MeV     & $775.8$  MeV \\
\hline
$m_{a_1}$  & $1230 \pm 40$  MeV   & $1220$  MeV      & $1230.6$  MeV  \\
\hline
$m_{\pi}$  & $139.6 \pm 0.0004$  MeV   & $141$  MeV     & $139.6$  MeV  \\
\hline
$f_{\pi}$  & $92.4 \pm 0.35$  MeV  & $84.0$  MeV     & $84.6$  MeV  \\
\hline
\end{tabular}  \\
\caption{meson masses without and with the chiral condensate.}
\end{center}
\end{table}

Finally, to confirm and understand numerical meson spectra in the $\s \to 0$ limit, we will
investigate \eq{eq:vector} and \eq{eq:axial} analytically in the small $\s$ case.
Notice that when the chiral condensate is very small, $F(z)$ and $\ph(z)$ can be
approximated as $1$ and $m_q z$ in the range of $0 \le z \le z_{IR}$.
Using this, the solution of the last equation in \eq{eq:axial} becomes
\be \la{relpc}
\pi(z) = \frac{m_{\pi}^2}{2 \pi m_q^2} \chi(z) + c,
\ee
where $c$ is a constant. From the boundary conditions, $\pi(0)=\chi(0)=0$, $c$ should
be fixed as $0$. Inserting this result into the second equation in \eq{eq:axial},
we finally obtain
\be
0 = \pa_z \ls \frac{1}{z} \pa_z \chi \rs +
       \frac{ m_{\pi}^2 - g_5^2 m_q^2 }{z} \chi .
\ee
From this, we can see that in the small chiral condensate limit
with very small quark mass, the pion mass is proportional to that of
the longitudinal mode of the axial vector meson $\chi$, which makes the pion mass
grow as the chiral condensate goes to zero.
In the $\s \to 0$ limit, the other equations for the vector and axial vector meson
can be rewritten as
\bea
0 &=& \pa_z \ls \frac{1}{z} \pa_z V^{(n)}  \rs + \frac{m_{\r}^2 }{z}
V^{(n)}   ,\nn
0 &=& \pa_z \ls \frac{1}{z} \pa_z \bar{A}_{\m} \rs + \frac{m_{a_1}^2 - g_5^2 m_q^2}{z}
\bar{A}_{\m} .
\eea
Therefore, we can see that the transverse mode of the axial vector $\bar{A}_{\m}$
and pion $\pi$ have the same mass at
$\s=0$, which was shown in the Table 2. Moreover, from the above we can find
a mass relation between the vector meson and pion
\be
m_{\r}^2 \approx m_{\pi}^2 - g_5^2 m_q^2 ,
\ee
which is not the exact relation but the approximation. So, in the case $m_q \ll m_{\pi}$,
we can see that the vector, axial vector meson and pion have the similar mass in the case
$\s=0$, which gives consistent results with Table 2. In the chiral limit, as shown in
the previous section, the pion mass becomes very small as $\s \to 0$. Though the chiral
symmetry is restored for $\s = 0$, the pion and the longitudinal mode of the axial vector
do not mix because the scalar field is zero in the chiral limit. So we can see that
the existence of the light quark mass significantly affects on the pion mass, see
Table 1 and 2.

\begin{figure}
\vspace{0cm}
\begin{center}
   \epsfxsize=7.5cm
   \epsfbox{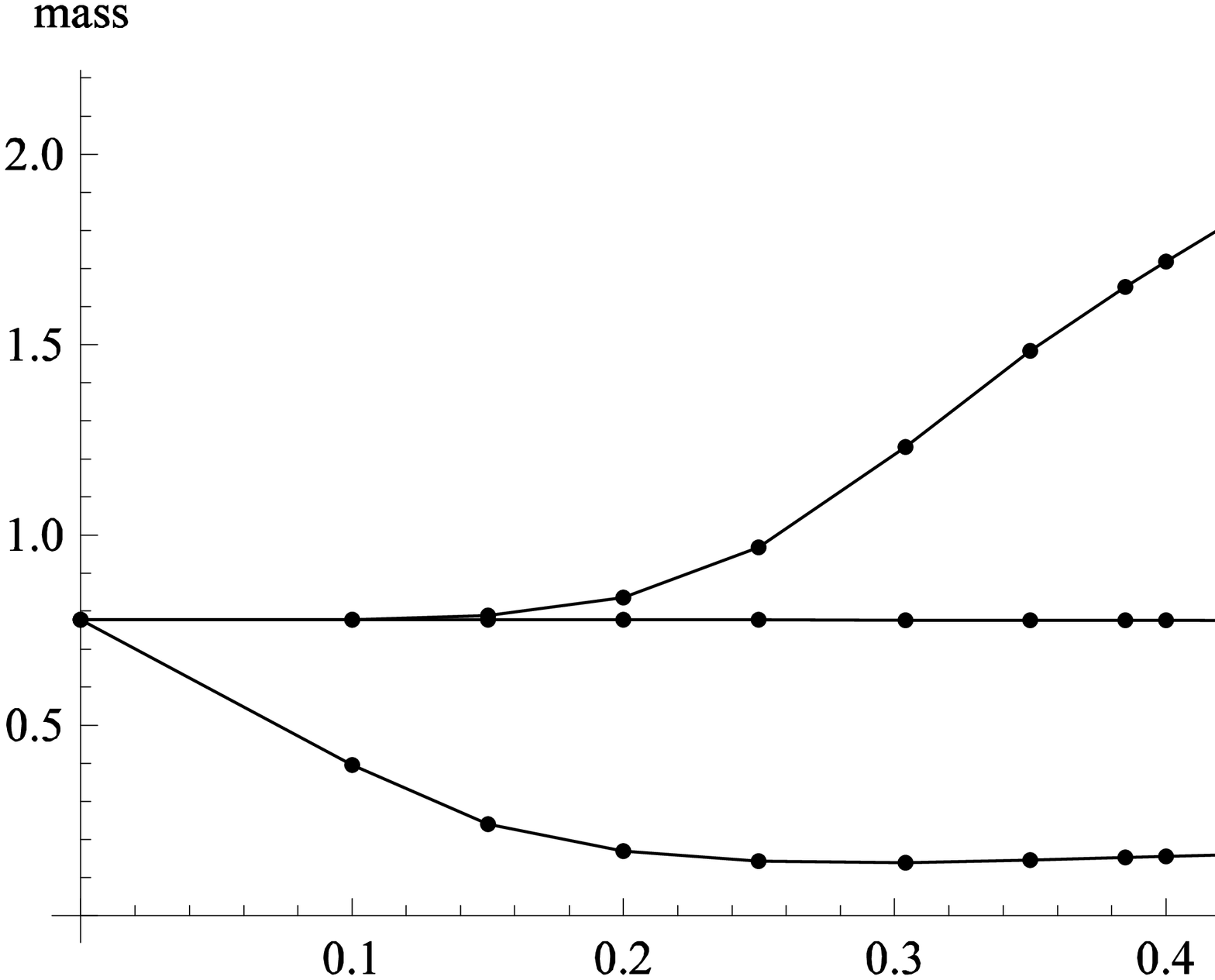}
   \vspace{0cm} \\
 $
  \begin{array}{ccc}
  \epsfxsize=7.5cm
   \epsfbox{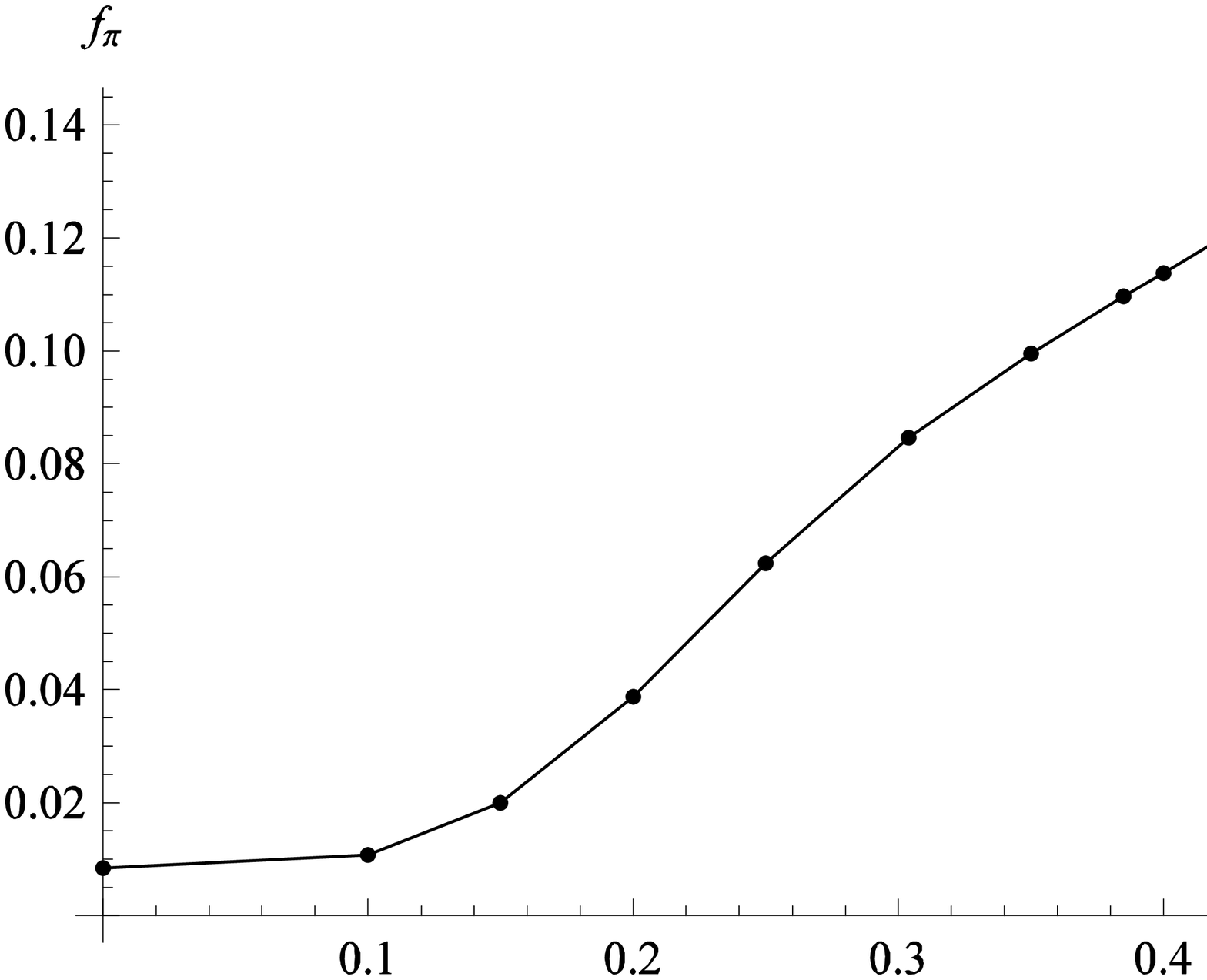} &
                           &
  \epsfxsize=7.5cm
   \epsfbox{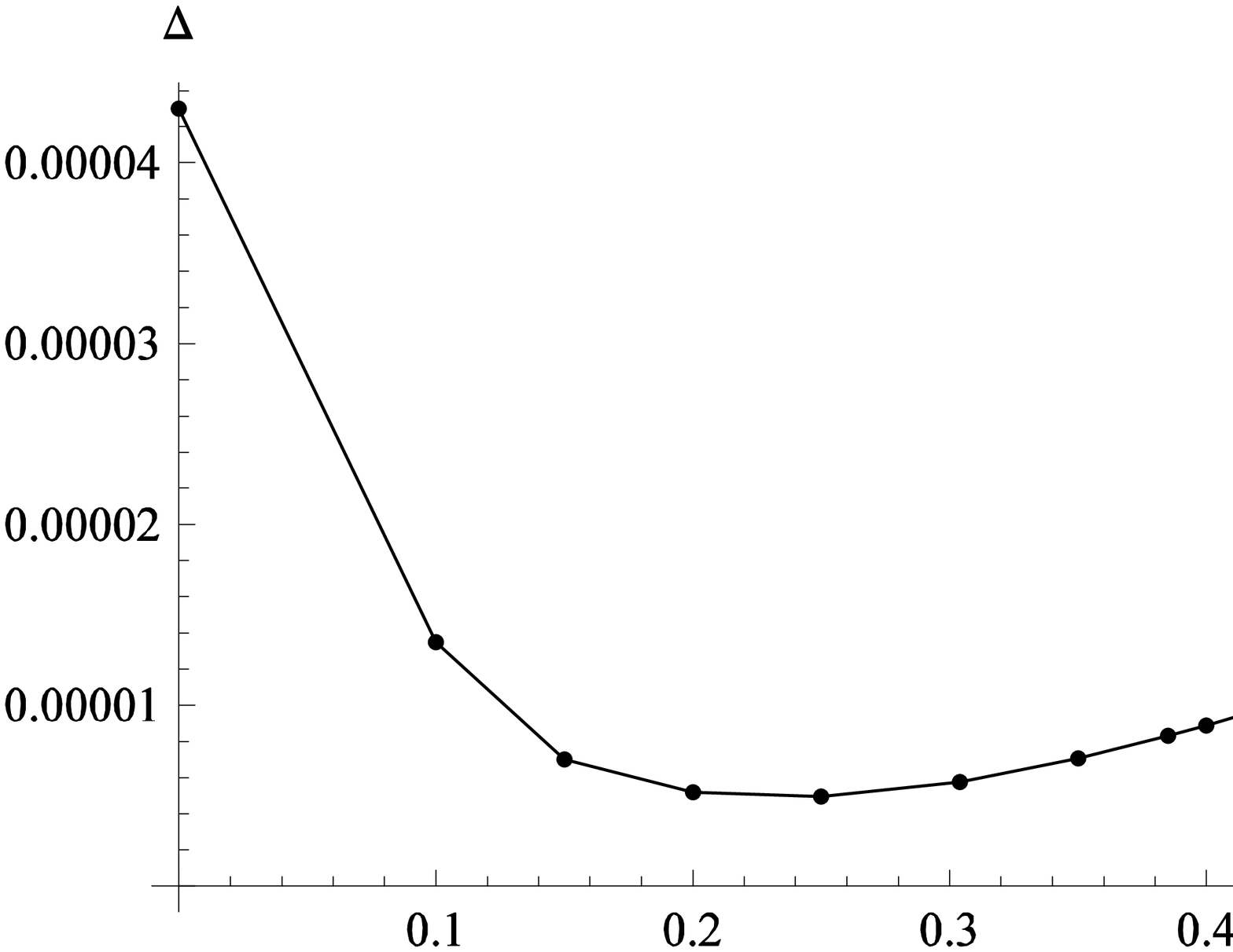}
  \end{array}
 $
 \vspace{-2.5cm}
\caption{For $m_q=2.383$MeV, we plot the masses of vector, axial vector meson and pion (the upper figure) and
the pion decay constant (left below). In the right below figure, the deviation from GOR relation
is plotted. For small chiral condensate regime, it seems that the GOR relation does not work. }
\label{non0mq_f_delta}
\end{center}
\end{figure}

\subsection{binding energy of the heavy quarkonium}

In the previous section, we have investigated the string breaking of the heavy quarkonium
when $m_q = 0$. In this section, we will study the string breaking for $m_q \ne 0$.
Since the method to calculate the interaction energy of the heavy quarkonium is the same
as the previous section, we summarize the results with some comments.

For $m_q \ne 0$, the interaction energy of the heavy quarkonium is given by the
same form as \eq{intenergy} with $F$ defined in \eq{fmqnzero}.
As mentioned, for the string breaking the interaction
energy should be greater than masses of two light quarks. In this case, since the light quark
has mass $m_q = 2.383$MeV the string breaking occurs at $V = 2 m_q$.

As we expect, since the quark mass is too small the string breaking distance is not
significantly changed from one obtained in the chiral limit, see Table 4.
Anyway, the non-zero quark mass make the string breaking distance large for the small
chiral condensate regime, roughly $\s \lesssim (0.45 {\rm GeV})^3$. This implies that
the non-zero light quark mass makes the string breaking slightly
more difficult.

\begin{table}
\begin{center}
 \begin{tabular}{|c|c|c|}
 \hline
 $\sigma$(GeV$^3$)  & distance for $m_q=0$ (GeV$^{-1}$) & distance for $m_q\neq0$ (GeV$^{-1}$)\\
 \hline\hline
  $(0.000)^3$ & 2.22426 & 2.25761  \\
 \hline
  $(0.100)^3$ & 2.22426 & 2.25760  \\
 \hline
  $(0.200)^3$ & 2.22406 & 2.25739  \\
 \hline
  $(0.250)^3$ & 2.22348 & 2.25681  \\
 \hline
  $(0.304)^3$ & 2.22171 & 2.25503  \\
 \hline
  $(0.350)^3$ & 2.21825 & 2.25155  \\
 \hline
  $(0.385)^3$ & 2.21342 & 2.24670 \\
 \hline
  $(0.400)^3$ & 2.21048 & 2.24375  \\
 \hline
  $(0.450)^3$ & 2.19467 & 2.22782  \\
 \hline
  $(0.500)^3$ & 2.12580 & 2.05779 \\
 \hline
 \end{tabular}
 \caption{String breaking distance depending on the chiral condensate in cases, $m_q = 0$ and $m_q \ne 0$.}
\end{center}
\end{table}


\section{Discussion}

We have studied $1/N_c$ correction of the meson spectra, which comes from the
light quark mass and/or chiral condensate. To encoding the chiral condensate
effect to the dual gravity theory, the gravitational backreaction of the massive
scalar field, which corresponds to light quark and/or the chiral condensate, was considered.
From the initial data
determined by the asymptotic solutions, we found the numerical solutions for metric
and scalar field, in which there exists a geometrical singularity. Because this
singularity causes the IR divergence of the dual gauge theory, we introduce an IR cut-off
to avoid it.

In this hard wall model, both the light meson spectra and heavy quarkonium
binding energy depending on the chiral condensate have been investigated.
For masses of the vector and axial vector meson, the chiral condensate dependence is almost
same in both cases $m_q = 0$ and $m_q \ne 0$. In other words, as the chiral condensate
increases, the vector meson mass decreases slightly while the axial vector meson mass increases
significantly.
In the $\s \to 0$ limit, since the chiral symmetry is fully restored the masses of
the vector and axial vector have the same value in the chiral limit. In the case of
$m_q \ne 0$, since the explicit light quark mass breaks the chiral symmetry slightly,
there exists small difference between the vector and axial vector meson masses.
As a result, the vector and axial vector meson masses do not significantly depend on
the light quark mass, while the pion mass depends on the light quark mass significantly.
In the chiral limit, due to the absence of the quark mass the pion mass is very
small compared with the observation. This implies that though light quark mass
is very small it plays a crucial role to determine the pion mass.
So we have recalculated the pion mass after encoding the light
quark mass to the dual background. Especially, we found that
due to this $1/N_c$ correction, in the case of $m_q = 2.383$ MeV and $\s = ( 304 {\rm MeV})^3$,
the original results obtained in \cite{Erlich:2005qh}
are improved to the real observations.

In the chiral limit, the pion mass increases as the chiral condensate increases, which
looks qualitatively consistent with the chiral perturbation theory in QCD.
However, in the case $m_q \ne 0$ there exists
a critical value of the chiral condensate $\s_c \approx   (0.26\mathrm{GeV})^3$.
As the chiral condensate grows, the pion mass decreases (increases)
below (above) this critical value. In the $\s \to 0$ limit for $m_q \ne 0$, the
pion and axial vector meson masses have the same value due to the mixing of them
as well as the restoration of the chiral symmetry. In the chiral limit,
though the chiral symmetry is fully restored,
there is no mixing between pion and axial vector meson in the $\s \to 0$ limit,
so that pion can have very small mass compared with the axial vector meson mass.
This unexpected behavior of the pion mass for $m_q \ne 0$ can be well understood
in the gravity side
but in the gauge theory side it is not clear why this behavior occurs.
The existence of the critical chiral condensate seems to indicate that the chiral perturbation
theory in QCD is only applicable above this critical value and below it
new approach is needed. So it would be very interesting
to understand, in the gauge theory side, the unexpected behavior of the pion mass below the critical
chiral condensate.

We also observed the heavy quarkonium binding energy from the string
breaking distance for zero and non-zero quark masses. On both cases, we found that
the string breaking distance becomes short as the chiral condensate increases.
This implies that the heavy quarkonium is broken to two heavy-light mesons
more easily at the large chiral condensate regime. We also investigated the string breaking
depending on the light quark mass. As we can expected, we found that since
the light quark mass is too smaller than the heavy quarkonium mass, the string breaking of the heavy
quarkonium  does not crucially depend on the light quark mass.

\vspace{1cm}

{\bf Acknowledgement}

C. Park thanks to Sang-Jin Sin and Youngman Kim for helpful discussion.
This work was supported by the National Research Foundation of Korea(NRF) grant funded by
the Korea government(MEST) through the Center for Quantum Spacetime(CQUeST) of Sogang
University with grant number 2005-0049409. C. Park was also
supported by Basic Science Research Program through the
National Research Foundation of Korea(NRF) funded by the Ministry of
Education, Science and Technology(2010-0022369).

\vspace{1cm}


\end{document}